\documentclass[twocolumn,english,aps,prb,amsmath,amssymb,superscriptaddress]{revtex4-1}
\usepackage[T1]{fontenc}
\usepackage[utf8]{inputenc}
\usepackage{xcolor}
\usepackage{babel}
\usepackage{amstext}
\usepackage{graphicx}
\PassOptionsToPackage{normalem}{ulem}
\usepackage{ulem}
\usepackage[unicode=true,
 bookmarks=true,bookmarksnumbered=false,bookmarksopen=false,
 breaklinks=false,pdfborder={0 0 0},backref=false,colorlinks=true]
 {hyperref}
\hypersetup{pdftitle={PTCDA Bilayer on Ag(111)},
 pdfauthor={NZ},
 citecolor=blue}
\usepackage{breakurl}

\makeatletter

\providecolor{lyxadded}{rgb}{0,0,1}
\providecolor{lyxdeleted}{rgb}{1,0,0}

\DeclareRobustCommand{\lyxdeleted}[3]{{\texorpdfstring{\color{lyxdeleted}\sout{#3}}{}}}

\@ifundefined{textcolor}{}
{%
 \definecolor{BLACK}{gray}{0}
 \definecolor{WHITE}{gray}{1}
 \definecolor{RED}{rgb}{1,0,0}
 \definecolor{GREEN}{rgb}{0,1,0}
 \definecolor{BLUE}{rgb}{0,0,1}
 \definecolor{CYAN}{cmyk}{1,0,0,0}
 \definecolor{MAGENTA}{cmyk}{0,1,0,0}
 \definecolor{YELLOW}{cmyk}{0,0,1,0}
}

\usepackage{babel}
\usepackage{siunitx}

\makeatother

\begin{document}

\title{Structure and vibrational properties of the PTCDA/Ag(111) interface:
Bilayer vs. monolayer}

\author{N. L. Zaitsev }

\email{nza@yandex.ru}

\affiliation{Laboratory of Theoretical Physics, Institute of Molecule and Crystal
Physics Ufa Research Center of Russian Academy of Sciences, 450075,
Ufa, Russia}

\affiliation{Laboratory of Nanostructured Surfaces and Coating, Tomsk State University,
634050, Tomsk, Russia}

\author{P. Jakob}

\affiliation{Department of Physics, Philipps-Universität Marburg, 35032, Marburg,
Germany}

\author{R. Tonner}

\affiliation{Department of Chemistry, Philipps-Universität Marburg, 35032, Marburg,
Germany}
\begin{abstract}
The structural and vibrational properties of metal-organic interfaces
have been examined by means of infrared (IR) absorption spectroscopy
and density functional theory (DFT) with an approach accounting for
long-range dispersive interactions. We focus on a comparative study
of the PTCDA monolayer and bilayer on Ag(111). The equilibrium geometry
at the molecule-metal interface and the infrared spectrum of the chemisorbed
monolayer of PTCDA on Ag(111) are well described by the computations.
In the bilayer structure, the presence of a physisorbed adlayer on
top of PTCDA/Ag(111) presents a challenge for DFT. As previously described
for other systems, the polarization of the substrate is not captured
correctly and results in too low energies of frontier molecular orbitals.
This results in an apparent contribution from the vibrations of second-layer
PTCDA to the IR spectrum from interfacial dynamical charge transfer
processes. After removing these peaks with artificially strong intensity,
calculated and experimental data show good agreement and the IR spectrum
can be described as the sum of the spectra of the PTCDA/Ag(111) contact
layer and a physisorbed PTCDA monolayer on top. 
\end{abstract}
\maketitle

\section{Introduction}

Organic molecular thin films are frequently used as integral parts
of micro- and optoelectronic devices, \cite{dimitrakopoulos_organic_2002,chidichimo_organic_2010,dou_lowbandgap_2015},
where interfaces between self-assembled planar organic molecules and
metallic substrate are important functional elements. Understanding
the physicochemical properties of metal-organic interfaces is essential
for the successful engineering of such devices. Processes taking place
within the first adsorbed organic layers largely determine the behavior
of the entire system \cite{kahn_electronic_2003,rusu_first-principles_2010}.
A comprehensive investigation of the organic contact layer and the
influence of the next adlayer will thus provide deeper insight into
organic materials on surfaces with well controlled properties.

Monolayers of PTCDA molecules on various silver surfaces have been
extensively examined using a wide range of experimental techniques.
\cite{tautz_2007_structure,willenbockel_interplay_2014} Thus, a detailed
description of adsorption geometry \cite{kraft_2006_lateral,hauschild_2005_molecular,hauschild_2010_normalincidence,kilian_molecular_2004,kilian_role_2008}
and electronic structure is available in the literature \cite{temirov_free-electron-like_2006,gerlach_substrate-dependent_2007,schwalb_electron_2008,yang_two-photon_2008,koch_energy_2008}.
Density functional theory (DFT) was repeatedly applied for a successful
description of PTCDA/Ag(111) as well \cite{rohlfing_adsorption_2007,romaner_theoretical_2009,rusu_first-principles_2010,zaitsev_change_2010,dyer_nature_2010,zaitsev_transformation_2012}.
This interface is often used as a benchmark for different approaches
to describe long range van der Waals (vdW) forces within DFT calculations
\cite{ruiz_2012_densityfunctional,maurer_2015_manybody}, which is
crucial for calculating accurate adsorption geometries. Thus, PTCDA/Ag(111)
is an excellent prototype system for exploring the influence of the
second layer on the properties of the first one.

Despite the vast opportunities of vibrational spectroscopy to characterize
organic layers on metal surfaces, very few studies on the vibrational
properties of PTCDA/Ag(111) exist \cite{tautz_strong_2002,wagner_raman_2003}.
In a recent study, the high intrinsic spectral resolution of infrared
absorption spectroscopy has been used to detect the vibrational line
shifts of only a few wavenumbers associated with different numbers
of layers of deposited molecular species, e.g. of bilayer, trilayer
or multilayer \cite{thussing_structural_2016,thussing_thermal_2017}.
The respective characteristic, layer-dependent vibrational modes can
be extracted from the experiment and compared to DFT calculations.
On one hand the experimental spectra are interpreted within a well
established ab initio framework, and on the other hand the accuracy
of DFT calculations are put to the test by precise experimental methods.
Note, that the reproduction of the vibrational properties of metal-organic
interfaces is a challenge for the current approximate exchange-correlation
functionals used for DFT computations insofar as the simultaneous
description of accurate adsorption geometry, energy level alignment
and the dipole layer formation are needed.

In the present study, we have examined PTCDA/Ag(111) by means of infrared
absorption spectroscopy and vdW-DFT calculations. The focus was on
a comparative study of the vibrational properties of the monolayer
and bilayer of PTCDA molecules on Ag(111). Here we confirm that already
a monolayer of organic molecules creates a metal-organic interface,
and the interface properties are hardly altered by increasing the
thickness of the organic adlayer.

\section{Methods}

Calculations were done using DFT as implemented in the SIESTA code
\cite{ordejon_1996_selfconsistent,soler_2002_thesiesta}. Localized
pseudoatomic orbitals were used for the wave function representation
together with norm-conserving pseudopotentials \cite{troullier_1991_efficient}.
Long-range dispersion forces were described by using the optB88-vdW
functional approach \cite{klimevs_2010_chemical,klimevs_2011_vander}
.

The scheme of periodically repeated slabs was used to describe the
Ag(111) surface. Interactions between the periodic images of the system
in the direction perpendicular to the surface ($z$-direction) was
suppressed by a vacuum layer of about 11~Å. The slab dipole correction
was used to avoid the artificial macroscopic electrostatic field which
arises due to the periodic boundary conditions \cite{neugebauer_1992_adsorbatesubstrate,bengtsson_1999_dipolecorrection}.
A uniform mesh for the numerical integration and solution of the Poisson
equation was specified by the energy cutoff of 400 Ry. The molecular
adsorbate was applied to one side of the substrate only. The surface
Brillouin zone sampling was done by 8 k-points with use of the Monkhorst-Pack
scheme. Atomic positions were optimized until the forces were smaller
than $\SI{0.01}{\electronvolt\per\angstrom}$. During the relaxation
the substrate was represented by four layers of atoms with the bottom
two layers fixed to bulk position.

The double-$\zeta$ polarized (DZP) basis set with an energy shift
of 10 meV and generated by a soft confinement scheme was used for
all atoms in the molecule. Two types of basis functions for silver
atoms were used. External atoms of the slab in the upper- and lowermost
silver layers have a cutoff radius $r_{c}=9.73$~a.u. of the $5s$
orbitals (\emph{energy shift} is 10 meV); for the other atoms and
the remaining internal Ag atoms $r_{c}=7.03$~a.u. has been used
(\emph{energy shift} is 180 meV). The same parameters were used and
justified in our recent works \cite{jakob_2016_adsorption,zaitsev_2016_adsorption}.
The silver lattice constant was set to its equilibrium value of $a=\SI{4.17}{\angstrom}$,
as was found from the calculation with the short $r_{c}$ basis functions.

Vibrational properties were calculated via diagonalization of the
dynamical matrix within the harmonic approximation. Finite displacements
of $\pm\SI{0.02}{\angstrom}$ of individual atoms in each spatial
direction were used to build the dynamical or force-constant matrix.
The sum rule which follows from Newton’s third law was imposed onto
the force constant matrix to ensure that the force acting on the dynamic
atom to be exactly the sum of the forces acting on the rest atoms
in the unit cell \cite{ackland_1997_practical}. The detailed description
of the procedure we were following here to get the vibrational frequencies
can be found in Ref. \cite{frederiksen_2007_inelastic}.

The vibrational analysis was carried out in the frozen surface approximation.
Only atoms of the PTCDA molecules were displaced for construction
of the Hessian \cite{li_2002_partial}, and Ag atoms were kept fixed
at their equilibrium positions. This is motivated by the large difference
in atomic mass between the light atoms of the adsorbate and the heavy
silver atoms. Infrared (IR) intensities were calculated as the square
of dipole moment derivatives with respect to the normal-mode coordinate
\cite{porezag_1996_infrared}. Because of the periodicity of the system
in $xy$-directions and the metallic nature of the Ag substrate, only
the $z$-component of the dipole moment was obtained in the calculation.
To visualize the IR spectrum, every vibrational mode was broadened
by Lorentzian ($\frac{A}{\pi}\frac{\sigma}{x^{2}+\sigma^{2}}$) with
$\sigma=\SI{2}{\per\centi\meter}$ and the corresponding amplitude
equals to the calculated IR intensity\@.

\section{Results}

It is well established,\cite{glockler_1998_highlyordered,kraft_2006_lateral}
that PTCDA molecules form an ordered monolayer on Ag(111). A herringbone
arrangement with two molecules per unit cell occupying nonequivalent
adsorption sites has been found {[}Fig.~\ref{fig:ads-dist-color}~(a){]}.
More recently, the adsorption geometry of PTCDA/Ag(111) was measured
with high accuracy using NIXSW \cite{hauschild_2005_molecular,hauschild_2010_normalincidence}
and calculated structures (vdW-DFT) were found to be in good agreement
with the experimental ones \cite{ruiz_2012_densityfunctional,maurer_2015_manybody}.

\begin{figure}
\includegraphics[width=0.9\columnwidth]{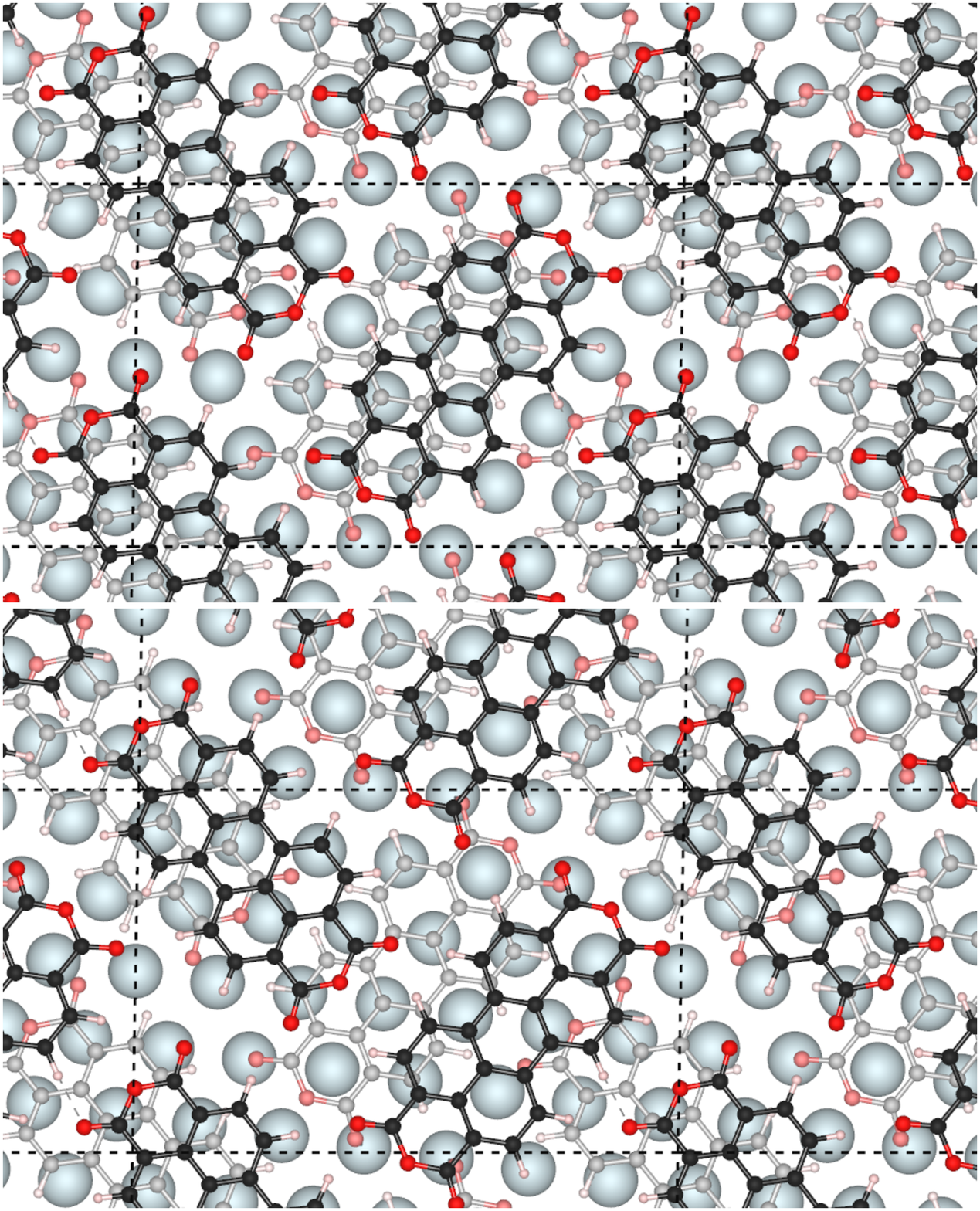}

\protect\protect\caption{\label{fig:Stacking} Considered arrangements of PTCDA bilayer structures
on Ag(111). (top) arrangement equivalent to the $\alpha$-phase \cite{ogawa_1999_34910perylenetetracarboxylic}
of the molecular crystal ($\alpha$-stacking) and (bottom) arrangement
equivalent to the $\gamma$-phase. The surface unit cell is denoted
by the black dashed line. The contact molecular layer is shown by
faded colors.}
\end{figure}

\subsection{Adsorption geometry}

\begin{figure*}
\includegraphics[width=0.95\textwidth]{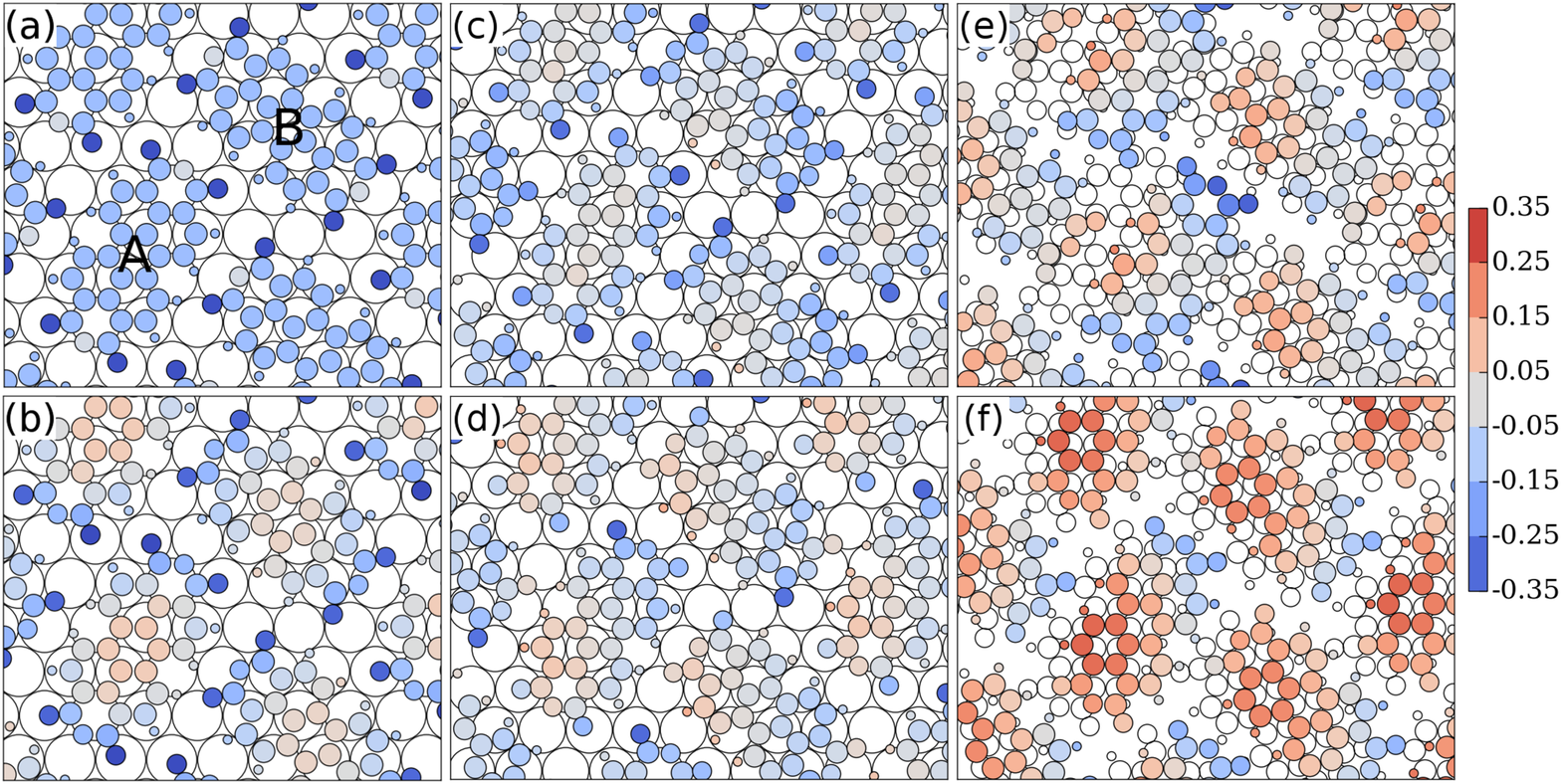}

\protect\protect\caption{\label{fig:ads-dist-color} Adsorption geometries of the PTCDA/Ag(111)
interface. The color code denotes deviation of the vertical adsorption
distance (in \AA ) from planes at $\SI{3}{\angstrom}$ (first adsorbed
layer) and $\SI{6}{\angstrom}$ (second adlayer) above the idealized
position of the silver topmost layer (extrapolated bulk positions).
The various systems presented in (a-f) are as follows: (a) experimental
data of the PTCDA/Ag(111) contact layer \cite{hauschild_2010_normalincidence},
(b) calculated optimized geometry of the PTCDA/Ag(111) monolayer,
(c) optimized position of first-layer PTCDA (P) in $\gamma$-P/P/Ag(111),
(d) in $\alpha$-P/P/Ag(111), (e) calculated optimized position of
second-layer PTCDA in $\gamma$-P/P/Ag(111) and (f) in $\alpha$-P/P/Ag(111).}
\end{figure*}

The optB88-vdW exchange-correlation functional used here has provided
equilibrium adsorption geometries of the PTCDA/Ag(111) monolayer in
reasonable agreement with the experimental measurements \cite{hauschild_2010_normalincidence,bauer_2012_roleof}.
The molecules are bent along the long axis; thereby, the anhydride
oxygen atoms reside higher from the substrate than the acylic ones,
with adsorption distances of $\sim\SI{2.7}{\angstrom}$ which is quite
close to the experimental value \cite{hauschild_2010_normalincidence}
of $\SI[separate-uncertainty=true]{2.66(3)}{\angstrom}$ {[}see fig.~\ref{fig:ads-dist-color}~(a,~b){]}.
Like in many vdW-DFT calculations,\cite{ruiz_2012_densityfunctional,bauer_2012_roleof,maurer_2015_manybody}
the carbon core adsorption distance of $\sim\SI{3.05}{\angstrom}$
is noticeably overestimated in comparison with experiment $\SI[separate-uncertainty=true]{2.86(1)}{\angstrom}$,\cite{hauschild_2010_normalincidence}
whereas the anhydride oxygen atom's adsorption height of $\sim\SI{2.85}{\angstrom}$
underestimates the experimental value of $\SI[separate-uncertainty=true]{2.98(8)}{\angstrom}$.
Our previous calculations of the similar system of NTCDA/Ag(111) performed
with the same method and parametrization \cite{jakob_2016_adsorption}
gave a similar picture of the adsorption geometry.

Based on experimental data, it was proposed \cite{kilian_molecular_2004,thussing_thermal_2017}
that PTCDA molecules in the top layer of the bilayer structure have
the same lateral order as in the contact layer but the relative positions
of the atoms in the top layer could not be identified. We thus checked
different layer stackings in the bilayer structure {[}denoted as P/P/Ag(111){]}.
The first structure represents the stacking of the layers in the $\alpha$-modification
of the molecular crystal of PTCDA\cite{ogawa_1999_34910perylenetetracarboxylic}
(denoted as $\alpha$-P/P/Ag(111) or $\alpha$-stacking). The second
structure results from shifting the top layer laterally (denoted as
$\gamma$-stacking to avoid confusion with $\beta$-modification of
the PTCDA molecular crystal\cite{ogawa_1999_34910perylenetetracarboxylic}).
We found the energy difference between the structural arrangements
to be negligible ($<\SI{0.01}{\electronvolt\per cell}$).

The adsorption height of the contact layer is noticeably altered due
to interaction with the second adlayer. This is apparent in Fig.~\ref{fig:ads-dist-color},
where the deviation of vertical adsorption distances from a plane
positioned $\SI{3}{\angstrom}$ above the extrapolated bulk position
of topmost silver layer of P/P/Ag(111) is depicted. Regarding $\gamma$-stacked
layers, the carbon atoms of the perylene core get closer to the substrate
by about $\SI{0.1}{\angstrom}$ {[}fig.~\ref{fig:ads-dist-color}~(b,~c){]},
whereas some acylic oxygen atoms move away by $\sim\SI{0.2}{\angstrom}$.
Even stronger distortions of the first layer are found for $\alpha$-P/P/Ag(111).

The second layer in both structures is distorted as well. As is clearly
visible in figure~\ref{fig:ads-dist-color}~(e,~f) the atomic positions
in this layer deviate considerably from the planar geometry for both
stacking types. Thereby the averaged interlayer distance in the $\alpha$-stacked
bilayer is larger by $\sim\SI{0.1}{\angstrom}$ as compared to $\gamma$-stacked
layers.

\subsection{Vibrational properties and infrared spectra}

Based on the obtained equilibrium structures, we performed calculations
of force-constant matrices for further analysis of the vibrational
properties and infrared spectra.

\subsubsection{PTCDA/Ag(111) Monolayer}

\begin{figure*}
\includegraphics[width=0.95\textwidth]{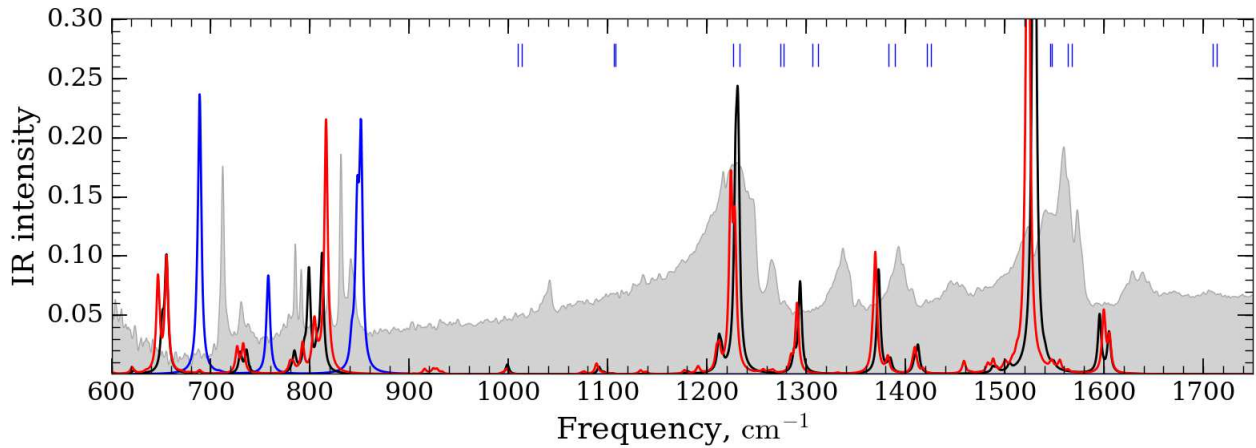}\protect\caption{\label{fig:IRAS}The measured (shaded area) and various calculated
vibrational spectra of a PTCDA monolayer: (i) free-standing layer
(blue); (ii) PTCDA/Ag(111) contact layer (black); (iii) $\gamma$-P/P/Ag(111)
bilayer with frozen second adlayer (red). An artificial line broadening
of $\SI{2}{\per\centi\meter}$ has been applied to the calculated
spectra\@. The intensities of the (calculated) modes with frequencies
below $\SI{900}{\per\centi\meter}$ were increased by a factor of
3\@. }
\end{figure*}

In the range of 600--900 $\si{\per\centi\meter}$ there are three
groups of prominent peaks in the experimental IR spectrum (shaded
area in fig.~\ref{fig:IRAS}) and they are well reproduced by the
calculated vibrational modes of PTCDA/Ag(111) (black curve), although
they display significantly shifted frequencies (about $\SI{60}{\per\centi\meter}$)
with respect to the experimental values. For planar molecular adsorbates
with their $\pi$-conjugated backbone oriented parallel to the surface
the spectral range above and below 900 $\si{\per\centi\meter}$ is
dominated by in-plane and out-of-plane modes, respectively.

Interestingly, the vibrational bands at 600--900 $\si{\per\centi\meter}$
consist of two splitted peaks. These spectrally close bands are well
reproduced by the calculation and they can be associated with the
two inequivalent molecules A and B interacting differently with the
Ag(111) substrate. This becomes more obvious when comparing the calculated
IR spectra of a free-floating PTCDA monolayer (blue curve) to PTCDA
adsorbed on Ag(111). Specifically, the contact to Ag(111) manifests
itself in a frequency downshift by about $\SI{40}{\per\centi\meter}$
and an increased peak splitting for each group. Since vibrational
modes in this region are IR-active because of their out-of-plane movement
of adsorbate atoms, the increased peaks separation most likely is
a consequence of the unequal adsorbate-substrate interaction.

The in-plane vibrational modes in the region between 1000 and 1700
$\si{\per\centi\meter}$ are quite prominent in the PTCDA/Ag(111)
experimental and calculated spectra (black line, fig. \ref{fig:IRAS}),
while they are completely absent in the IR spectrum of the free PTCDA
monolayer (blue line). Apparently, they appear only in the presence
of a metallic substrate; this is because the vibrational dipole moment
is enhanced due to interfacial dynamic charge transfer (IDCT) between
the metal and the molecular layer \cite{persson_vibrational_1980,braatz_2012_vibrational,rosenow_2016_electrontextendashvibron,tonner_molecular_2016}.

When visualizing the oscillatory motion of atoms within the PTCDA
molecule, the individual symmetry of each mode can be derived. In
the case of in-plane modes of PTCDA all but a few weak bands at 1091,
1181 and 1389 $\si{\per\centi\meter}$ belong to the $\textrm{A}_{1}$
irreducible representation of the $\textrm{C}_{2\textrm{v}}$ -- symmetry
group. The mentioned weak features belong to the $\textrm{A}_{2}$
irreducible representation; they become dipole active due to a local
symmetry reduction $\textrm{C}_{2\textrm{v}}\rightarrow\textrm{C}_{2}$
as a consequence of neighboring PTCDA and the herringbone arrangement
of PTCDA/Ag(111).

As in the case of out-of-plane modes, the overall shape of the experimental
spectrum at $\omega>\SI{1000}{\per\centi\meter}$ is well reproduced
by the calculated spectrum of the PTCDA/Ag(111) contact layer. In
order to assess the effect of the Ag(111) substrate on the vibrational
mode frequencies we have connected the line positions of related modes
for adsorbed PTCDA and the free-floating PTCDA monolayer by thin lines.\lyxdeleted{nz}{Tue Jan 23 06:27:51 2018}{
} Note that the broad bands in the experimental spectrum at about
1230 and 1540 $\si{\per\centi\meter}$ are severely influenced by
anharmonic coupling of weak combination/overtone bands with nearby
strong fundamental modes ($\rightarrow$ Fermi resonance coupling)
and associated intensity transfer \cite{jakob_fermi_1998,braatz_2012_vibrational,breuer_vibrational_2012}.
Such couplings cannot be captured with our current approach of calculating
vibrational mode frequencies and intensities, i.e. reproduction of
the sub-structure of these bands is not possible. This extra complexity
in the vibrational spectrum is particularly obvious in the 1500--1580
$\si{\per\centi\meter}$ region where the calculation yields a single
intense band at 1530 $\si{\per\centi\meter}$, corresponding to a
C-C bond stretching vibration along the longitudinal axis of the molecule.
This shows the limit of the currently employed approximations.

\begin{figure*}
\includegraphics[clip,width=0.9\textwidth]{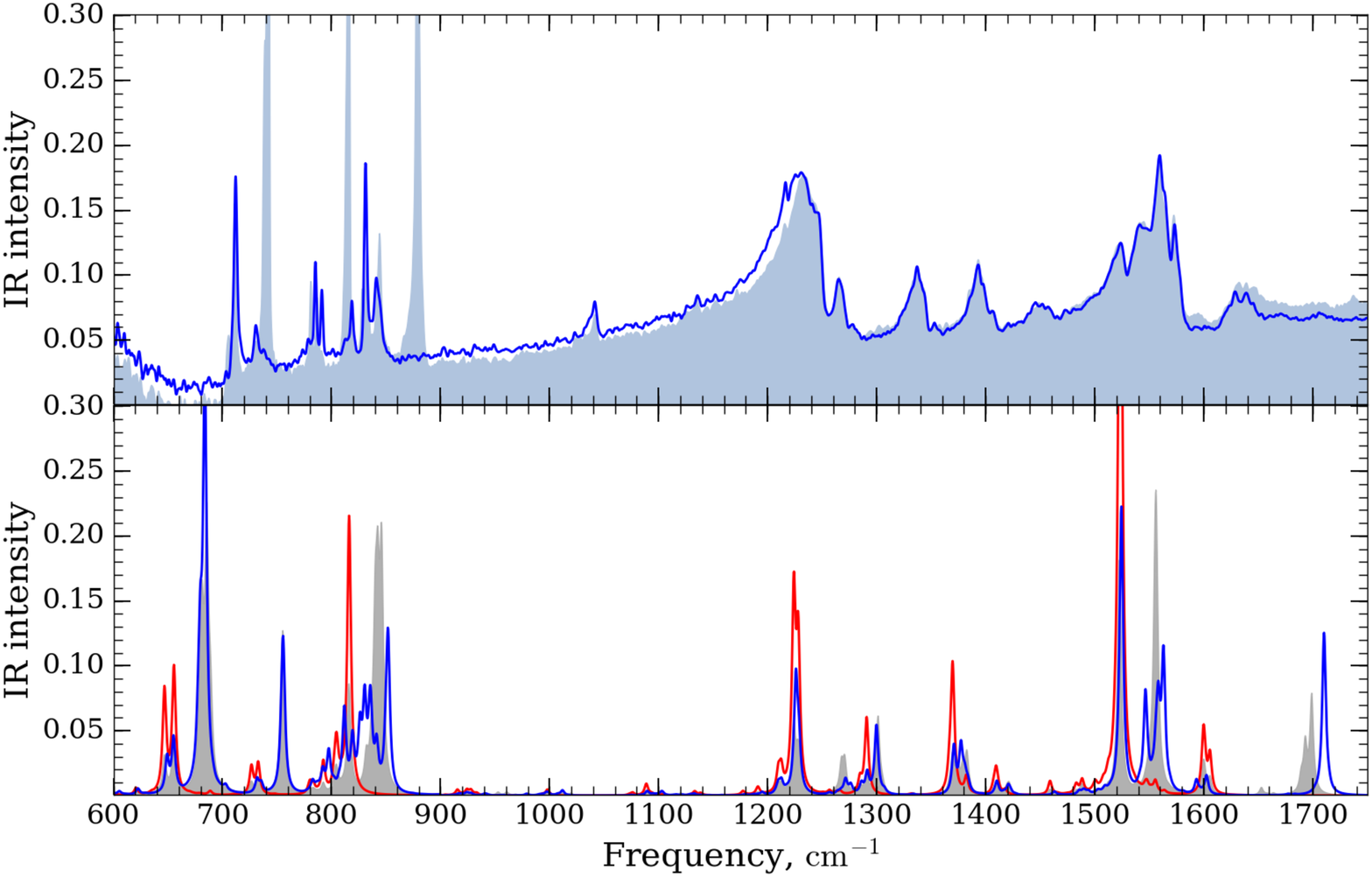}

\protect\caption{\label{fig:IRAS_Bilayer}(top) Experimental IR absorption spectra
for the PTCDA/Ag(111) monolayer (blue line) and the PTCDA/Ag(111)
bilayer (shaded area). (bottom) Calculated vibrational spectra of
various $\gamma$-P/P/Ag(111) bilayers: (i) full spectrum (shaded
gray); (ii) spectrum with frozen second adlayer (red); (iii) spectrum
with planarized second layer (blue), i.e. with all internal bonds
oriented parallel to the surface. An artificial line broadening of
$\SI{2}{\per\centi\meter}$ has been applied to the calculated spectra\@.
The intensities of the (calculated) modes with frequencies below $\SI{900}{\per\centi\meter}$
were increased by a factor of 3\@.}
\end{figure*}

The two closely spaced bands at $\sim\SI{1630}{\per\centi\meter}$
deserve special mentioning; they correspond to stretching modes of
the C-O$_{acyl}$ bonds of the molecules A and B (fig.~\ref{fig:IRAS})
and, upon adsorption, they are more strongly shifted to lower frequencies
as compared to other modes. This was already previously ascribed to
the covalent interaction of acyl oxygen atoms with nearby Ag atoms
of the substrate \cite{rosenow_2016_electrontextendashvibron}.

\subsubsection{PTCDA/Ag(111) Bilayer}

A comparison of experimental spectra associated with the mono- and
bilayers of PTCDA on Ag(111) shows that the effect of the second adlayer
on the vibrational spectrum primarily concerns the out-of-plane modes
at $<\SI{1000}{\per\centi\meter}$ and is minor at $>\SI{1000}{\per\centi\meter}$
(fig. \ref{fig:IRAS_Bilayer}); in-plane modes of second-layer PTCDA
apparently exhibit zero or near-zero dynamic dipole moments. This
means that interfacial dynamical charge transfer (IDCT), which is
the main cause of the pronounced intensity of first-layer PTCDA in-plane
modes, is not operative for second-layer PTCDA.

\begin{figure*}[t]
\includegraphics[width=0.9\textwidth]{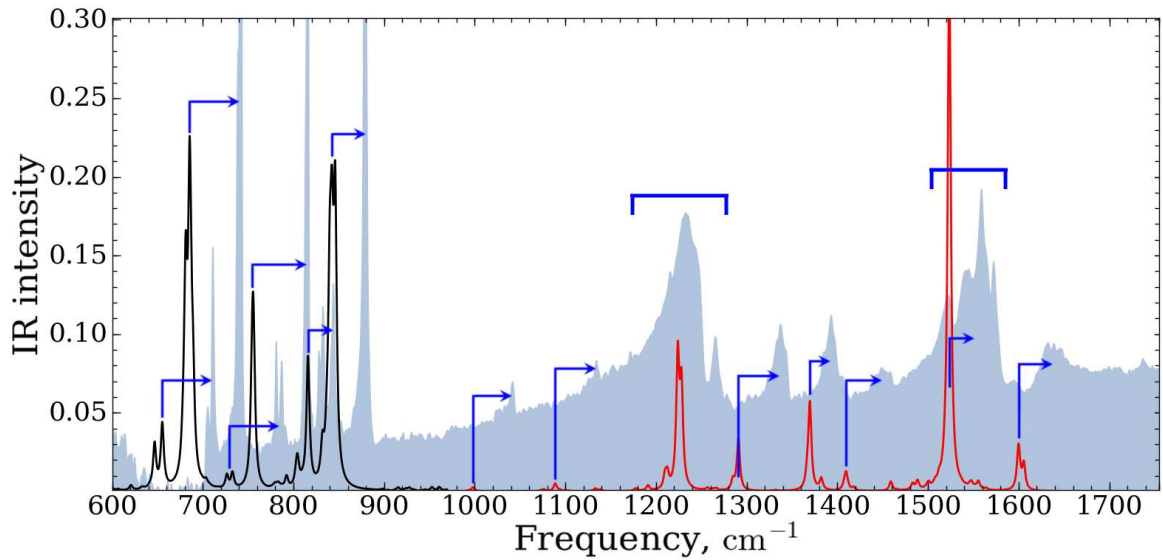}

\protect\caption{\label{fig:Bilayer_Reproduction}Reproduction of the experimental
PTCDA/Ag(111) bilayer spectrum using a combination of calculated spectra.
The low frequency region below $\SI{900}{\per\centi\meter}$ is reproduced
by the full spectrum of $\gamma$-P/P/Ag(111); at $\omega>\SI{900}{\per\centi\meter}$
a frozen second layer is used to eliminate spurious bands associated
with second-layer vibrations. Blue arrows have been added to indicate
the frequency deviation between calculated and experimental value.
An artificial line broadening of $\SI{2}{\per\centi\meter}$ has been
applied to the calculated spectra\@. The intensities of the (calculated)
modes with frequencies below $\SI{900}{\per\centi\meter}$ were increased
by a factor of 3\@. }
\end{figure*}

The red curve in fig.~\ref{fig:IRAS} refers to the IR spectrum of
the PTCDA/Ag(111) bilayer, calculated with the second layer being
pinned to its equilibrium geometry (i.e. the force-constant matrix
contains only atomic displacements in the first layer). Despite noticeable
differences in adsorbate-substrate vertical separation of PTCDA/Ag(111)
in comparison with P/P/Ag(111) {[}fig.~\ref{fig:ads-dist-color}~(b,~c,~d){]}
the line positions of the various modes of the PTCDA contact layer
are changed insignificantly (see fig.~\ref{fig:IRAS}), i.e. the
peaks are shifted by few wave numbers only. This clearly demonstrates
that the mere presence of a second PTCDA adlayer has a distinct, but
only slight effect on the full range of calculated IR spectrum of
the PTCDA contact layer. Furthermore, according to the red line and
gray shaded area on fig.~\ref{fig:IRAS_Bilayer} this also holds
for vibrations of first-layer PTCDA when the motion of atoms in both
layers are explicitly included into the dynamical matrix.

In the region of out-of-plane vibrational modes three additional peaks
(fig.~\ref{fig:IRAS_Bilayer}) emerge in the calculated full spectrum
of $\gamma$-P/P/Ag(111), in accordance with the measurements, their
frequencies are shifted towards the blue. Their line positions are
very similar to those of the free-floating PTCDA monolayer, which
indicates that second-layer PTCDA is interacting only weakly with
the underlying molecular layer as well as the Ag(111) substrate. This
conclusion is corroborated by the narrow line shapes (and accordingly
significantly enhanced peak intensities) of the respective bands in
fig.~\ref{fig:IRAS_Bilayer}.

Surprisingly, a fair number of frequency-shifted new peaks appear
in the calculated full spectrum at $\omega>\SI{1000}{\per\centi\meter}$,
i.e. in the spectral region of in-plane vibrational modes (gray shaded
area, fig.~\ref{fig:IRAS_Bilayer}). Comparison to fig.~\ref{fig:IRAS}
suggests that these additional features stem from vibrations of the
topmost adlayer. These shifts are quite small $\sim\SI{10}{\per\centi\meter}$
for the two central peaks at $\sim\SI{1300}{\per\centi\meter}$ and
$\SI{1380}{\per\centi\meter}$, while somewhat larger values ($\sim\SI{30}{\per\centi\meter}$)
are found for the prominent band at $\sim\SI{1520}{\per\centi\meter}$
and the weak feature at $\SI{1230}{\per\centi\meter}$ (see fig.~\ref{fig:IRAS_Bilayer}).
A particularly pronounced shift is observed for the C-O$_{acyl}$
stretching mode (located at about 1600 $\si{\per\centi\meter}$ for
first-layer PTCDA) which is located at $\sim\SI{100}{\per\centi\meter}$
higher frequencies for second-layer species. 

The vibrational frequencies and their patterns of movement are virtually
the same for the top-layer of PTCDA and a free PTCDA monolayer sheet.
In striking contradiction to the experimental observations, however,
the dynamic dipole moments of these new bands is substantial for the
top-layer. Note that their intensity cannot be ascribed to slight
deformations of the planar structure or a possible tilt, as a very
similar spectral signature is found for a perfectly flat second adlayer
(fig.~\ref{fig:IRAS_Bilayer}). 

\begin{figure}
\includegraphics[width=1\columnwidth]{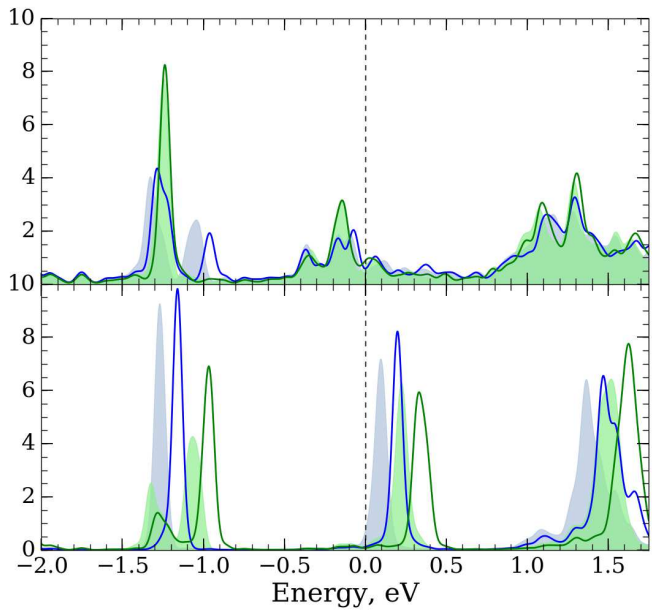}

\protect\protect\caption{\label{fig:pdos} Density of states projected on $p_{z}$-orbitals
of each PTCDA molecule in the unit cell of $\gamma$-P/P/Ag(111).
Contact (\emph{first}) molecular layer (top panel) and topmost (\emph{second})
molecular layer (bottom panel). Molecule of type A (B) is indicated
in blue (green) color {[}see also fig.~\ref{fig:ads-dist-color}~(a){]}.
Density of states calculated with slab dipole correction \cite{neugebauer_1992_adsorbatesubstrate,bengtsson_1999_dipolecorrection}
is shown by a solid line, otherwise by shaded areas. The Fermi energy
is indicated by a dashed vertical line.}
\end{figure}

As discussed above (and more thoroughly in the literature \cite{braatz_2012_vibrational,tonner_molecular_2016,rosenow_2016_electrontextendashvibron})
the only way that in-plane vibrational modes of parallel adsorbed
PTCDA may become IR-active is IDCT. We note that IDCT associated with
the excitation of vibrational modes implies a partial filling of the
lowest unoccupied molecular orbital (LUMO) of PTCDA molecules in the
top layer. According to the calculated density of states, the LUMO
of these species is located a mere $\SI{0.2}{\electronvolt}$ above
the Fermi level (see fig.~\ref{fig:pdos}), whereas experiment suggests
notably higher energies, which renders it entirely unoccupied \cite{yang_two-photon_2008}.

It was established \cite{neaton_2006_renormalization,garcia-lastra_2009_polarizationinduced}
that weak coupling between a metal surface and adsorbed molecules
results in a polarization of the substrate, that has a large influence
on the energy level alignment. This nonlocal correlation effect can
not be described properly within common DFT calculations; rather,
a renormalization of Kohn-Sham molecular electronic levels is necessary
\cite{egger_2015_reliable,ma_2016_theenergy}.

In our case, the second adlayer is weakly coupled to the contact layer
on the metal substrate and it is likely that the DFT calculation predicts
an incorrect alignment of the levels of the molecular adlayer with
respect to the Fermi energy of the Ag substrate. Indeed, in contrast
to the chemisorbed monolayer, unoccupied levels of weekly coupled
molecules in the second adlayer are sensitive to slab polarization
by an external electric field which exactly compensates the artificial
dipole moment in slab calculations \cite{neugebauer_1992_adsorbatesubstrate,bengtsson_1999_dipolecorrection}.
As a consequence the LUMOs of molecules A and B in the second layer
shift by $\sim\SI{0.1}{\electronvolt}$ towards higher energy as clearly
seen in figure~\ref{fig:pdos}, while the partly filled former LUMOs
of PTCDA molecules in the contact layer keep their energies. Therefore
aforementioned renormalization of the molecular levels (that has not
explicitly be carried out in this study) will move the LUMO of second-layer
PTCDA molecules higher in energy, consistent with the experimental
values. Thereby, apparent IDCT effects between Ag(111) and the top-layer
molecules are eliminated.

If this artificial IDCT effect is removed, all contributions from
the in-plane vibrational modes of the molecules of the topmost organic
layer vanish in the calculated IR-spectrum, in accordance with the
measurements. At the same time, out-of-plane vibrational modes in
the range of 600--950 $\si{\per\centi\meter}$ are subject to only
minor changes. A reasonable description of the PTCDA bilayer IR spectrum
in the entire spectral range can thus be given by a superposition
of two spectra (i) of the PTCDA/Ag(111) contact layer and (ii) of
the free-floating PTCDA monolayer; alternatively, one may combine
the full spectrum of the PTCDA/Ag(111) bilayer (at $\omega<\SI{900}{\per\centi\meter}$)
and of the $\gamma$-P/P/Ag(111) spectrum with frozen second layer
(at $\omega>\SI{900}{\per\centi\meter}$), as has been done in fig.~\ref{fig:Bilayer_Reproduction}.

\begin{figure}
\includegraphics[clip,width=1\columnwidth]{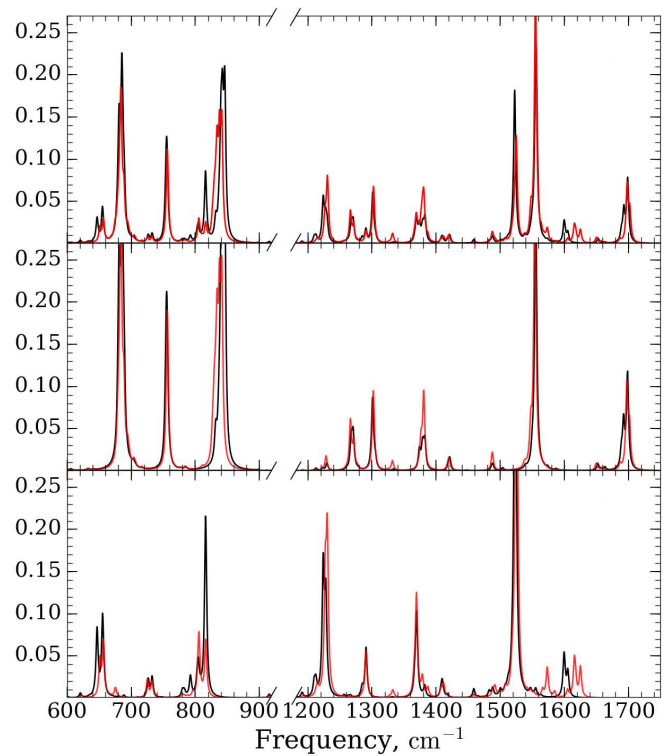}

\protect\caption{\label{fig:alpha_gamma}Calculated IR-spectra of $\gamma$-P/P/Ag(111)
(black line) and $\alpha$-P/P/Ag(111) (red line). Full spectrum (top
panel), spectrum with frozen \emph{contact} layer (in the middle)
and spectrum with frozen \emph{second} adlayer (bottom panel). An
artificial line broadening of $\SI{2}{\per\centi\meter}$ has been
applied to the calculated spectra\@. The intensities of the (calculated)
modes with frequencies below $\SI{900}{\per\centi\meter}$ were increased
by a factor of 3\@.}
\end{figure}

It is worthwhile to note that despite considerably different adsorption
geometries of $\gamma$- and $\alpha$-stacked bilayers on Ag(111)
only small differences are found in the calculated IR spectra; this
in particular holds for the vibrational bands at $\omega<\SI{900}{\per\centi\meter}$
which display an only weak selectivity regarding $\gamma$ and $\alpha$
discrimination (see Fig.~\ref{fig:alpha_gamma} top panel). Worth
mentioning are the enhanced splitting of first-layer PTCDA modes at
about $\SI{650}{\per\centi\meter}$ and the multiple bands at $\SI{780}-\SI{820}{\per\centi\meter}$
which experience an intensity gain on the high frequency side for
the $\gamma$-stacked bilayer, in accordance with the experiment (see
fig.~\ref{fig:IRAS_Bilayer}).

Upon inspection of the in-plane vibrational modes we find that both,
$\gamma$- and $\alpha$-stacked bilayers suffer from the erroneously
enhanced dynamic dipole moments of second-layer PTCDA species which
add extra features to the calculated spectra in the top and middle
panels of fig.~\ref{fig:alpha_gamma} (spectral region $\omega>\SI{900}{\per\centi\meter}$).
This is why evaluation of the bottom panel (where the influence of
the two stacking schemes on first-layer molecules is shown) is more
useful. Most features at $\omega>\SI{900}{\per\centi\meter}$ are
virtually identical for $\gamma$ and $\alpha$-stacked bilayers;
the only exception is the characteristic C-O stretching mode of first-layer
PTCDA which is shifted by $\sim\SI{20}{\per\centi\meter}$ (fig.~\ref{fig:alpha_gamma})
for the $\alpha$-stacked bilayer. This is at variance with the experiment
(fig.~\ref{fig:IRAS_Bilayer}), and an additional argument that the
PTCDA/Ag(111) bilayer adheres to $\gamma$ stacking.

\section{Summary}

The influence of the second adlayer on the structural and vibrational
properties of the PTCDA/Ag(111) contact layer was studied by infrared
adsorption spectroscopy in combination with density functional theory
calculations. The case of strong covalent interaction between a monolayer
of PTCDA molecules and the Ag(111) substrate is quite well described
by the vdW-DFT calculations, which provide a good agreement with the
experimental adsorption geometry as previously shown. In addition,
the calculated fundamental vibrational modes, obtained in the harmonic
approximation, together with IR-intensities derived from the dynamic
dipole approach give a reasonable description of the measured IR-spectrum.

The second adlayer affects the equilibrium positions of first-layer
molecules quite noticeably, and the adsorption height depends strongly
on the layer arrangement. The change in adsorption geometry reflects
itself in the IR-spectrum; especially the C-O stretching and C-H out-of-plane
bending modes react sensitively, while the other ones show only small
changes.

The comparison of experimental IR spectra with the computations represents
a sensitive test of the weak interaction between second-layer PTCDA
and the PTCDA/Ag(111) contact layer. The accurate experimental data
can help here to reveal an improper energy level alignment at the
interface in the DFT calculations as is well-known for many exchange
correlation functionals. This misalignment results in artificial interface
dynamical charge transfer between Ag substrate and the top-layer molecules,
which in turn creates spurious bands in the IR spectrum, in contradiction
with the experiment.

Clearing of the calculated IR-spectrum from these erroneous contributions
provides a good agreement with the measurements.
\begin{acknowledgments}
This work was supported by the Deutsche Forschungsgemeinschaft (DFG)
within SFB 1083 ``Structure and Dynamics of Internal Interfaces''.
Computational resources were provided by HRZ Marburg, HLR Stuttgart
and CSC-LOEWE Frankfurt. 
\end{acknowledgments}

 \bibliographystyle{apsrev4-1}
\bibliography{ref}

\end{document}